\documentclass[prd,showpacs,showkeys,amsmath,amssymb,10pt]{revtex4}
\usepackage{amsfonts}
\usepackage{graphicx}
\usepackage{dcolumn}
\usepackage{bm}

\def\text#1{\mbox{#1}}

\def\ee{\end{equation}}
\def\be#1{\begin{equation}\label{#1}}

\def\ba{\begin{array}}
\def\ea{\end{array}}

\begin{document}
\title{Generalized axion-photon couplings and dual descriptions on brane-worlds}
\author{M. O. Tahim}
\email{mktahim@fisica.ufc.br}
\author{C. A. S. Almeida}
\email{carlos@fisica.ufc.br} \affiliation{Departamento de
F\'{\i}sica, Universidade Federal do Cear\'a, Caixa Postal 6030,
60455-760, Fortaleza, Cear\'a, Brazil}

\begin{abstract}
In this work we obtain topological and dual theories on
brane-worlds in several dimensions. Our brane is a solitonic-like
hypersurface embedded in a space-time with a specific
dimensionality and it appears due to the breaking of a
Peccei-Quinn-like symmetry. In the first part of this work, the
obtained topological theories are related to a generalization of
the axion-foton anomalous interaction in $D=4$ (in the Abelian
case) and to the Wess-Zumino term (in the non-Abelian case). In
the second part, we construct dual models on the brane through a
mechanism of explicit Lorentz symmetry breaking. The gauge
symmetries of such models are discussed within the Stuckelberg
formalism.
\end{abstract}


 \vspace{1.0cm}

\maketitle

\section {Introduction}
In current brane-world research the most important subject has to do
with (fermionic or bosonic) matter field localization procedures.
Its importance resides in the fact that is just this procedure that
will give physical information about the universe inside the brane.
Such information are interaction energy scales, the range of the
interactions, the chiralities of the fields, symmetries, etc. In
this direction, hard work have been made. In the case of gravity
localization procedures, the standard way commonly accepted is that
given by the Randall-Sundrum scenario \cite{RS}. This scenario is
characterized by treating Einstein gravity in classical level.
However, there are a lot of approaches and interpretations to
Einstein gravity, one of them is the so called $B\wedge F$-Gravity
\cite{bf-gravity}. Such formalism, strictly related to Loop Gravity,
is known by providing a consistent machinery of quantization of
gravity. An interesting peculiarity of this type of theory is that
it is completely background independent \cite{no_metric}.

Due to these developments, we regard as an important subject look
for topological theories in the brane context, with the major
objective to add quantum information to the Randall-Sundrum
scenario. Pursuing this idea, as the first step, we study
topological and equivalent theories on brane-worlds in several
dimensions and discuss some mechanisms of dimensional reduction,
namely, the naive one and a mechanism involving branes. In the
first part of this work we construct topological theories in
brane-worlds. The brane-world is regarded as a kink-like soliton
\cite{dw_brane} and it appears due to a spontaneous symmetry
breaking of a Peccei-Quinn-like symmetry. Topological terms are
obtained by generalizing to several dimensions the axion-foton
anomalous interaction (in the Abelian case) and the Wess-Zumino
term (in non-Abelian theories). Indeed, this anomalous interaction
is valid in a grand unification scenario with the electroweak
angle $sin^{2}\theta_{\omega}=\frac{3}{8}$. Such effect is mainly
due to the presence of instantons in the Standard Model
\cite{huang,kymieong}. However the generalizations made in this
work are simply inspired in this 4D anomalous interaction and they
aim in to preserve the field theories basic symmetries in order to
open the possibilities of application of these new topological
terms in brane-world scenarios. Despite the fact that such
theories do not have propagating local degrees of freedom, they
are important in dual descriptions, regardless of other
applications.

In the second part, we construct dual tensor gauge theories on the
brane. Investigations of dual formulations for tensor fields are
important for understanding of alternative formulations of known
theories like gravity as well as understanding of their role in
superstrings. Also, such models are important because, in the
non-Abelian case, they describe a low energy region of a QCD
system and, since that the topological properties do not depend on
small distances (local characteristics) but depend on the global
ones, this has a direct application in describing the IR regimen
of QCD, where the perturbation theory fails down \cite{duality}.
Besides this, we point out the fact that these models are
equivalent to free and propagating bosonic theories, an
interesting characteristic that may be used in the context of
localization scenarios. As we will show, this construction
involves Lorentz symmetry breaking. We also discuss, at the end,
the gauge symmetries of these models within the Stuckelberg
formalism. In this case we note the breaking of a gauge symmetry
of the type $U(1)\times U(1)$ down to $U(1)$.

\section {Topological theories on domain wall-branes (Abelian
case)}

In what follows we use capital letters to denote space-time
indices in a $D>4$ space-time, greek indices stand for a $D=4$
space-time and lower cases (a, b, c, d...) stand for a $D<4$
space-time. We implement the theory through the following action
in $D=5+1$:
\begin{equation}
S=\int d^{6}x\left( -\frac{1}{2\left( 3!\right) }H_{MNP
}H^{MNP}+g\varepsilon ^{MNPQRS}\phi
\left( z\right) H_{MNP}H_{QRS}+\frac{1}{2}%
\partial _{M}\phi \partial ^{M }\phi +V\left( \phi \right)
\right)  . \label{eq1}
\end{equation}
In this action, $H_{MNP}=\partial _{M}B_{NP
}+\partial _{N }B_{PM}+\partial _{P }B_{MN
}\left( M,...,S =0,...,5\right) $ is the field strength
of an antisymmetric tensor gauge field $B_{MN }$. The field
$B_{MN}$ has an important function in string theory: it
couples correctly with the string world-sheet in a very similar
way to the coupling of a gauge vector field $A_{M }$ to the
universe line of a point particle. The field $\phi $ is a real
scalar field, and $V\left( \phi \right) $  is a potential that
provides a phase transition:
\begin{equation}
V\left( \phi\right) =\lambda \left( 1-\cos \phi\right).
\label{eq2}
\end{equation}
In fact, the solution for the
specified potential, is
\begin{equation}
\phi(z)=\pi+2\arcsin(\tanh \sqrt{\lambda}z),
\end{equation}
for $\phi\in (0,2\pi)$ when $z\rightarrow \pm \infty$. This solution
corresponds to the topological sector which contains just one
soliton \cite{kymieong,sikivie}. The second term in the action
(\ref{eq1}) is a term that generalizes the coupling  that appears
from the anomaly of the Peccei-Quinn quasisymmetry
\cite{peccei-quinn} in $D=3+1$, namely, $\phi\rightarrow\phi+2\pi$.
This symmetry is broken by the potential term written with the real
scalar field $\phi$. In a phase transition, only the later field
acquires a nonzero vacuum expectation value (VEV). For the case of
the action (\ref{eq1}) the phase transition favors the creation of
domains of different phases: in fact, the vacuum is made of many
disconnected points, i. e., the potential is minimized when $\phi
_{vacuum}=2\pi n$, where \textbf{n} is a integer number. The domains
created are separated by dynamical solitonic hypersurfaces
topologically stables. In this sense, these objects are brane-like
objects \cite{dw_brane}. For such, the space-time dimension is
$D=5+1$ and the hypersurface is a $D=4+1$ world. Now we can work
with the second term of Eq.(\ref{eq1}) (considering that $\phi$ only
depends on the z-coordinate \cite{kymieong}) in order to obtain new
terms by integration by parts. The first term obtained is a total
derivative and may be disregarded. Another term is identically null
because of the antisymmetry of the Levi-Civita symbol. As the field
$\phi$ depends only on the z-coordinate, the topological term may be
rewritten as follows:
\begin{equation}
S_{top.}=3\alpha \int d^{5}xdz\left( \varepsilon ^{3NPQRS
}\partial _{3}\phi\left( z\right) B_{NP}H_{QRS}\right) .
\label{eq5}
\end{equation}
Considering that the $B_{MN}$ field weakly depends on the
z-coordinate, Eq.(\ref{eq5}) is rewritten as:
\begin{equation}
S_{top.}=\int d^{5}x\left( k\varepsilon ^{3NPQRS }B_{NP}H_{QRS
}\right). \label{eq6}
\end{equation}
This last equation shows that over the hypersurface an effective
topological term appears with a coupling constant \textbf{k} that
have canonical dimension of mass. This coupling constant is
quantized in various ways \cite{kymieong,jackiw}. The theory over
the hypersurface is completely five-dimensional. This term is very
similar to the Chern-Simons term \cite{jackiw}, which is written in
$D=2+1$ with a gauge vector field $A_{\mu }$:
\begin{equation}
S_{cs}=g\int d^{3}x\varepsilon ^{abc }A_{a }F_{bc}. \label{eq7}
\end{equation}
Nevertheless, the Eq.(\ref{eq6}) is written only with tensorial
antisymmetric fields $B_{MN}$. Such term have been used to explain
some peculiarities of the Cosmic Microwave Background Radiation
(CMBR) \cite{indianos} within the Randall-Sundrum scenario
\cite{RS}. It is interesting now to observe the properties of the
action (\ref{eq1}) in lower dimensional space-times using
dimensional reduction. Thus, supposing that the fields of the
action (\ref{eq1}) are independent of the coordinate $x_{M }\equiv
x_{5}$ which is not the argument of the field $\phi\left(
z\right)$ and defining
\[
B_{P 6}=V_{P },
\]
\[
B_{6P }=-V_{P },
\]
\[
V_{MN }=\partial _{M }V_{N}-\partial _{N }V_{M },
\]
\begin{equation}
\varepsilon ^{3NPQRS }\equiv \varepsilon ^{NPQRS} , \label{eq8}
\end{equation}
the action (\ref{eq1}) becomes:
\begin{equation}
S=\int d^{5}x\left( -\frac{1}{4}V_{MN }V^{MN }-\frac{1}{2\left(
3!\right) }H_{MNP }H^{MNP }+g\varepsilon ^{MNPQR}\phi \left(
z\right) V_{MN }H_{PQR }+\frac{1}{2}\partial _{M }\phi\partial ^{M
}\phi +V\left( \phi \right) \right) . \label{eq9}
\end{equation}
This action in $D=4+1$ has now a vectorial gauge field $V_{M}$
reminiscent of the reduction, and contains yet the real scalar
field $\phi$, that again may give rise to the formation of a lower
dimensional domain wall-brane. In this case, the space-time
dimension is $D=4+1$ and the hypersurface is a $D=3+1$ universe.
If we observe the theory over the solitonic hypersurface we will
obtain that, rewriting the topological term of the action
(\ref{eq9}) as made in Eqs.(\ref{eq5}), (\ref{eq6}) and
(\ref{eq7}):
\begin{equation}
S_{top.}=\int d^{4}x\left( k\varepsilon ^{4\nu \alpha \rho \sigma
}V_{\nu \alpha }B_{\rho \sigma }\right)  . \label{eq10}
\end{equation}
We can note that the theory on the domain wall is strictly
four-dimensional. If the field $V_{\mu}$ is identified with the
potential four-vector $A_{\mu }$ then we obtain the action for the
$B\wedge F$ model \cite{bf-models} on the domain wall-brane. This
action, under certain conditions, can give rise to a mechanism of
topological mass generation for the field $A_{\mu }$ or for the
field $B_{\mu \nu }$.

Starting from Eq.(\ref{eq9}), the discussion for lower dimensions
($D=3+1$ and $D=2+1$), using the same methods, will lead to the
following topological action:
\begin{equation}
S_{top.}=\int d^{4}xk\left[ \varepsilon ^{\mu \nu \alpha \rho
}\phi \left( z\right) \partial _{\mu }\varphi \partial _{\nu
}B_{\alpha \rho }+\varepsilon ^{\mu \nu \alpha \rho }\phi \left(
z\right) F_{\mu \nu }W_{\alpha \rho }\right] .  \label{eq12}
\end{equation}
The fields $\varphi $ and $W_{\alpha \rho }=\partial _{\alpha
}W_{\rho }-\partial _{\rho }W_{\alpha }$ emerge as degrees of
freedom reminiscent of the reduction. These fields are defined as
in Eq.(\ref{eq8}). If we work with the first term of
Eq.(\ref{eq12}) on the domain wall, we will find a different
topological theory \cite{dedit-bfi}, namely:
\begin{equation}
S=\int d^{3}x\left( g\varepsilon ^{abc }\partial _{a }\varphi
B_{bc }\right) .  \label{eq13}
\end{equation}
Identifying again in the second term of Eq.(\ref{eq12}), the vector
field $W_{\mu }$ as the gauge field $A_{\mu }$, we will obtain the
anomalous interaction term between the real scalar field $\phi$ and
the field $A_{\mu }$. This term, rearranged on the domain wall,
reduces to the Chern-Simons term given by Eq.(\ref{eq7}).

\section {Topological theories on domain wall-branes (non-Abelian
case)}

Non-Abelian theories provide alternative mechanisms of treating
and quantization of gravity \cite{jackiw}. Recent interest have
appeared in the study of metric independent gravity theories
\cite{no_metric}. As applications of the results discussed above,
we will show how to obtain non-Abelian topological terms on domain
wall-branes. These type of terms are important for studies about
topological gravity trapped on brane-worlds.

In this case, the brane emerges from the same mechanism discussed
above. In our description, we consider the following
action:
\begin{equation}
S=\int d^{4}x\left( \frac{1}{2}\partial _{\mu }\phi \partial ^{\mu
}\phi +V\left( \phi \right) +k\varepsilon ^{\mu \nu \alpha \rho
}\phi \left( z\right) F_{\mu \nu }^{i}F_{\alpha \rho }^{i}\right).
\label{eq14}
\end{equation}
In this action, $F_{\mu \nu }^{i}=\partial _{\mu }A_{\nu
}^{i}-\partial _{\nu }A_{\mu }^{i}+gf^{ijk}A_{\mu }^{j}A_{\nu
}^{k}$, where $i,j,k=1,...,n$ and $\phi$ is a real scalar field.
We consider $i,j,k,l...$ as indices of a finite dimensional
semi-simple Lie group $G$. As already discussed, writing the third
term of Eq.(\ref{eq14}) over the hypersurface, we obtain the
following effective action:
\begin{equation}
S=k\int d^{3}x\varepsilon ^{\mu \nu \alpha }\left( \partial _{\mu
}A_{\nu }^{i}A_{\alpha }^{i}+gf^{ijk}A_{\mu }^{i}A_{\nu
}^{j}A_{\alpha }^{k}\right) .\label{eq15}
\end{equation}
This term is rather similar to the non-Abelian Chern-Simons term.
As discussed by Deser and Jackiw \cite{jackiw}, the non-Abelian
Chern-Simons term is not invariant under large gauge
transformations. However, this behavior is avoided if we consider
quantization of the coupling constant of this theory. The term
found in Eq.(\ref{eq15}) may be used to describe gravity in $D=3$
in the same way as in the references \cite{no_metric,jackiw}.

Another model may be obtained starting from the following action
in $D=5$:
\begin{equation}
S=\int d^{5}x\left( \frac{1}{2}\partial _{M}\phi\partial ^{M }\phi
+V\left( \phi \right) +k\varepsilon ^{MNPQR} \phi\left( z\right)
H_{MNP}^{i}F_{QR}^{i}\right) , \label{eq16}
\end{equation}
where $H_{MNP}^i\equiv \partial _M  B_{NP}^i+
\partial _N  B_{PM}^i+\partial _P B_{MN}^i+
g'f^{ijk}A_M ^j B_{NP }^k$ $\left( M,N,... =0,...,5\right)$. In
this last case, after simple calculations we obtain an action that
contains an effective topological term of the non-Abelian $B\wedge
F$ type:
\begin{equation}
S=k\int d^{4}x\varepsilon ^{\mu \nu \alpha \rho }B_{\mu \nu
}^{i}F_{\alpha \rho }^{i} . \label{eq17}
\end{equation}
This action is regarded as basic for treating metric independent
gravity theories in $D=4$ as topological constrained field theories
\cite{bf-gravity}.

\section {Dual theories on domain wall-branes}

In this section we discuss a procedure for construction of dual
theories on domain wall-branes. We start with the following action
in $D=5$:
\begin{equation} \label{eq18}
S=\int d^{5}x[\frac{1}{2}\partial_{M}\phi\partial^{M}\phi-V(\phi)-
\frac{k}{2}\varepsilon^{MNPQR}\phi\partial_{M}B_{NP}F_{QR}-\frac{k'}{2}B^{MN}B_{MN}\partial_{P}V^{P}].
\end{equation}
The field $\phi$ yet provides the emergence of the domain
wall-brane, in this case, a $3$-brane. The field $F_{MN}$ is the
field strength for the gauge field $A_{M}$, i.e.,
$F_{MN}=\partial_{M}A_{N}-\partial_{N}A_{M}$ (we are treating the
Abelian case) and $B_{MN}$ is an antisymmetric tensor field, the
Kalb-Ramond field. The vector $V^{M}$ in the fourth term
represents an additional parameter of the theory, i.e., it
represents another gauge freedom of this theory. This freedom can
be fixed by choosing $V^{M}$ pointing along a preferential
direction. With a choice like this we can break the $SO(1,4)$
Lorentz symmetry of the model. Similar procedure have been made in
the context of topological gravity theories \cite{lee smolin}.
Another interesting attempts have been made in scenarios with
Lorentz symmetry breaking \cite{jackiw-manoel-messias}.

In the background of a domain wall-brane $\phi\equiv\phi(x_4)$,
and we choose $V^{\mu}=(0,0,0,0,\phi)$ in such a way that the last
terms of the action (\ref{eq18}) can be rearranged:
\begin{equation}\label{eq19}
S\sim \int
d^{5}x[\frac{k}{2}\varepsilon^{4NPQR}\partial_{4}\phi(x_{4})B_{NP}F_{QR}+
\frac{k'}{2}\partial_{4}\phi(x_{4})B^{MN}B_{MN}].
\end{equation}
We note that this last action is yet invariant under
$\phi\rightarrow\phi+2\pi$. Making a thin wall approximation, i.e.,
$\partial_{4}\phi(x_{4})=\delta(x_{4})$ and defining
$g^{2}=\frac{k'}{k}$ we obtain
\begin{equation}\label{eq20}
S\sim k\int dx_{4}\delta(x_{4})\int
d^{4}x[\frac{1}{2}\varepsilon^{\nu\alpha\beta\lambda}B_{\nu\alpha}F_{\beta\lambda}+\frac{1}{2}g^{2}B^{\alpha\beta}B_{\alpha\beta}],
\end{equation}
where we have made the identification
$\varepsilon^{4\nu\alpha\beta\lambda}\equiv\varepsilon^{\nu\alpha\beta\lambda}$.
The conclusion is that we obtain the $B\wedge F$-Maxwell model in
$D=4$, namely
\begin{equation}\label{eq21}
S\sim \int
d^{4}x[\frac{1}{2}\varepsilon^{\nu\alpha\beta\lambda}B_{\nu\alpha}F_{\beta\lambda}+
\frac{1}{2}g^{2}B^{\alpha\beta}B_{\alpha\beta}].
\end{equation}
This model, as it is well known, is equivalent to the free gauge
invariant (non-massive) Maxwell theory \cite{duality}. We arrive at
this result by explicitly breaking the $SO(1,4)$ Lorentz symmetry
down to $SO(1,3)$ through the $V^{\mu}$ choice and the supposition
$\phi\equiv\phi(x_4)$.

The procedure explained above can be applied to other sort of
theories. In particular, we can obtain similar results for lower
dimensional theories. For example, we consider the following
models in $D=4$:
\begin{equation}\label{eq22}
S=\int
d^{4}x[\frac{1}{2}\partial_{\mu}\phi\partial^{\mu}\phi-V(\phi)-
k\varepsilon^{\mu\nu\alpha\beta}\phi\partial_{\mu}W_{\nu}F_{\alpha\beta}+k'W^{\mu}W_{\mu}\partial_{\lambda}V^{\lambda}]
\end{equation}
and
\begin{equation}
S'=\int
d^{4}x[\frac{1}{2}\partial_{\mu}\phi\partial^{\mu}\phi-V(\phi)-
k\varepsilon^{\mu\nu\alpha\beta}\phi\partial_{\mu}B_{\nu\alpha}\partial_{\beta}\varphi-
\frac{k'}{2}B^{\alpha\beta}B_{\alpha\beta}\partial_{\lambda}V^{\lambda}].
\end{equation}
In these cases we have more fields: $W^{\mu}$ is an Abelian gauge
field in the first model and $\varphi$ is a real scalar field in the
second model. These two theories have $SO(1,3)$ Lorentz symmetry.
Again, choosing $V^{\mu}=(0,0,0,\phi)$ and $\phi\equiv\phi({x_3})$
we break $SO(1,3)$ down to $SO(1,2)$. In the thin wall approximation
we are lead to the following theories:
\begin{equation}\label{eq23}
S\sim \int d^{3}x
[\varepsilon^{abc}W_{a}F_{bc}-g^{2}W^{a}W_{a}]
\end{equation}
and
\begin{equation}
S'\sim \int d^{3}x
[\varepsilon^{abc}B_{ab}\partial_{c}\varphi+\frac{1}{2}g^{2}B^{ab}B_{ab}].
\end{equation}
The first model is the $B\wedge F$-Maxwell model in $D=3$ which is
dual to a free and non-massive Maxwell theory. The last model is
called $B\varphi$-Klein-Gordon model \cite{dedit-bfi} which is also
dual to the free and non-massive Maxwell theory or yet to the free
and non-massive Klein-Gordon theory.

It is interesting to study the gauge symmetries of these models.
It is clear, for example, that the action (\ref{eq18}) is not
invariant under the usual gauge transformations (regarding
$V_{M}$ as a gauge field):
\begin{equation}
\delta\phi=0 ,\quad \quad \delta A_{M}=\partial_{M}\alpha ,
\quad \quad \delta V_{M}=\partial_{M}\overline{\alpha}
,\quad \quad \delta B_{MN}=\partial_{[M}\Lambda_{N]}.
\label{eq24}
\end{equation}
In order to recover the gauge symmetry we can make use of the
Stuckelberg formalism \cite{stuckelberg} simply redefining the
fields of the model:
\begin{equation}\label{eq25}
A^{M}\rightarrow A^{M}+\frac{1}{g}\partial^{M}\theta, \quad
\quad V^{M}\rightarrow V^{M
}+\frac{1}{g'}\partial^{M}\overline{\theta}, \quad \quad
B_{MN}\rightarrow B_{MN}+\partial_{[M}\Gamma_{N]}.
\end{equation}
Defining the new gauge transformations
\begin{eqnarray}
\delta\phi=0 , \quad \quad \delta A_{M}=\partial_{M}\alpha ,
\quad \quad \delta\theta=-g\alpha, \quad \quad \delta
V_{M}=\partial_{M}\overline{\alpha}, \\ \nonumber
\delta\overline{\theta}=-g'\overline{\alpha}, \quad \quad \delta
B_{MN}=\partial_{[M}\Lambda_{N]}, \quad \quad \delta
\Gamma_{N}=-\Lambda_{N}  ,\label{eq26}
\end{eqnarray}
we recover the gauge symmetry of the model. An interesting
characteristic is that the action (\ref{eq18}) contains two gauge
fields, namely $A_{M}$ and $V_{M}$ and, therefore, the gauge
symmetry for the vector fields is of the type $U(1)\times U(1)$. A
symmetry like this have been discussed in the context of models for
superconductors \cite{domain-gauge}. The important point here is
that when we make the choice for the direction of the vector $V^{M}$
we also break the $U(1)\times U(1)$ gauge symmetry down to $U(1)$,
i.e., the Stuckelberg formalism can be applied in the domain
wall-brane to restore this $U(1)$ gauge symmetry. The same arguments
are valid for other sort of theories.

\section {Discussions and Outlook}

In this work we have discussed how to construct topological and dual
theories in domain wall-branes in several dimensions. In the first
part, we have discussed the appearance of several Chern-Simons-like
topological terms, in Abelian and in non-Abelian theories, by a
procedure of dimensional reduction (the naive one). We have seen
that the brane breaks the Lorentz symmetry of these models. In order
to do this, the domain wall-brane has been simulated by a kink-like
soliton embedded in a higher dimensional space-time and it has
emerged  due to a spontaneous symmetry breaking of a specific
discrete symmetry, namely, a Peccei-Quinn-like symmetry.

In the second part of this work, we have developed a way to built
dual models in domain wall-branes. We have constructed, from a $D=5$
theory, the $B\wedge F$-Maxwell model in $D=4$ and, from a $D=4$
theory, the $B\wedge F$-Maxwell model and the
$B\varphi$-Klein-Gordon both in $D=3$. The procedure adopted consist
in the explicit breaking of the Lorentz symmetry due to the choice
of a preferential direction for the additional vector parameter
$V^{\lambda}$ of the theories. Interestingly, in the case of the
$B\wedge F$-Maxwell model, we can implement through the Stuckelberg
formalism a $U(1)\times U(1)$ gauge symmetry that is broken down to
$U(1)$ in the domain wall-brane due to our choice. Another
characteristic is that we always obtain in the domain wall-brane,
theories that are dual to free and gauge invariant models. Indeed,
this fact is compatible with the idea of localization of fields in
brane worlds, where the zero modes (described by non-massive
theories) mimics the standard model fields. In that case, the
existence of bosonic modes is guaranteed if we assume the existence
of kinetic terms in the world-volume as well as in the bulk of the
brane. In branes which contain matter, this can be turned out by
perturbative corrections \cite{dvali}. On the other hand, we argue
in this work that those kinetic terms can be generated using tensor
like dualities. Hence its importance for mechanisms of field
localization in branes. A complete analysis of such idea should be
interesting. We consider also that the generalization of these dual
models for the non-Abelian case should be important. In this case,
applications to gravity theories would be possible.

The authors would like to thank Funda\c{c}\~{a}o Cearense de apoio
ao Desenvolvimento Cient\'{\i}fico e Tecnol\'{o}gico (FUNCAP) and
Conselho Nacional de Desenvolvimento Cient\'{\i}fico e
Tecnol\'{o}gico (CNPq) for financial support.

\end{document}